\documentclass[traditabstract]{aa} 

\usepackage{graphicx}
\usepackage{txfonts}
\usepackage{booktabs}
\usepackage{bm}
\usepackage{booktabs}

\newcommand{\msun}{\ensuremath{\mathrm{M}_{\odot}}\xspace}

\newcommand{\code}[1]{\texttt{#1}}
\newcommand{\mesa}{\code{MESA}\,}
\newcommand{\gyre}{\code{GYRE}\,}

\newcommand{\gdor}{\ensuremath{\gamma {\rm\,Dor}}\xspace}
\newcommand{\bv}{Brunt-V\"{a}is\"{a}l\"{a}\xspace}

\usepackage{xcolor}

\definecolor{green}{rgb}{0.3,0.7,0.}

\newcommand\new[1]{{#1}}

\begin{document}

\title{A method for non-linear inversion of the stellar structure applied to gravity-mode pulsators}
\author{Eoin Farrell\inst{1}, Gaël Buldgen\inst{1,2}, Georges Meynet\inst{1}, Patrick Eggenberger\inst{1}, Marc-Antoine Dupret\inst{2}, Dominic M. Bowman\inst{3, 4}}
\institute{Department of Astronomy, University of Geneva, Chemin Pegasi 51, 1290 Versoix, Switzerland
\and
STAR Institute, Université de Liège, Liège, Belgium
\and 
School of Mathematics, Statistics and Physics, Newcastle University, Newcastle upon Tyne, NE1 7RU, UK 
\and
Institute of Astronomy, KU Leuven, Celestijnenlaan 200D, Leuven, B-3001, Belgium 
}

\abstract
{We present a method for a non-linear asteroseismic inversion suitable for gravity-mode pulsators and apply it to slowly pulsating B-type (SPB) stars. Our inversion method is based on the iterative improvement of a parameterised static stellar structure model, which in turn is based on constraints from the observed oscillation periods. We present tests to demonstrate that the method is successful in recovering the properties of artificial targets both inside and outside the parameter space. We also present a test of our method on the well-studied SPB star KIC 7760680. We believe that this method is promising for carrying out detailed analyses of observations of SPB and \gdor stars and will provide complementary information to evolutionary models.}

\titlerunning{Non-linear inversions of SPB stars}
\authorrunning{Farrell et al.}


\maketitle

\section{Introduction}

The internal structure of stars during the core hydrogen burning phase is impacted by chemical mixing due to convection, rotation, and other physical processes. The nature of this internal mixing in intermediate and massive stars ($\gtrsim 2 \msun$) remains relatively unconstrained despite its importance in their subsequent evolution. Asteroseismology can help to provide further insight into the internal structure of these stars by reconstructing aspects of the stellar interior based on observed global oscillation modes. However, extracting useful information about the stellar structure from the observed oscillation modes is often quite challenging. To work towards tackling this challenge, we present a method for the non-linear inversion of the stellar structure applied to intermediate-mass main-sequence (MS)  stars that exhibit high-radial-order gravity modes.

Theoretical investigations of slowly pulsating B-type (SPB) stars and \gdor stars suggest that observed gravity modes can be used to constrain properties of the near-core region of the stellar interior \citep[e.g.][]{Miglio2008, Aerts2010, Aerts2019}.
In a classical study, \citet{Miglio2008} presented an analytical analysis of the impact of a spike in the \bv frequency caused by the chemical composition gradient outside a convective core. They show that mode trapping leads to periodic dips in the period spacing patterns with characteristics that related to the \bv profile. These studies also presented models with different assumptions for overshooting and diffusive mixing and discussed the impact on the period spacing patterns for \gdor and SPB stars.
More recent work has built on this in a variety of ways (see the reviews by \citealt{Aerts2019, Bowman2020}).
\citet{Pedersen2018} investigated the extent to which the shape of convective core overshooting and additional mixing in the envelope can be constrained using gravity-mode period spacings and concluded that gravity modes can be used to distinguish between step and exponential overshooting during the early part of the MS phase.
\citet{Wu2018} proposed the variation in period spacing as a signature of evolutionary status.
\citet{VanReeth2018} investigated the application of gravity modes to study differential rotation in intermediate-mass MS stars.
\citet{Michielsen2019} studied whether the thermal structure and shape of a near-core mixing profile can be distinguished through asteroseismology. 
More recently, \citet{Hatta2023} built on the results from \citet{Miglio2008} to develop semi-analytical expressions of period spacing patterns with different \bv profiles.

The use of gravity modes from SPB and \gdor stars to probe the convective core boundaries has been made possible in the last decade due to long-baseline observations from the CoRoT mission \citep{Auvergne2009, Baglin2009} and \textit{Kepler} mission \citep{Koch2010, Borucki2010}.
Uniform period spacing of gravity modes in a massive star were first reported by \citet{Degroote2010}.
Subsequent observations have been used to infer the presence of long series of consecutive high-radial-order dipole gravity modes in SPB stars such as KIC 10526294 \citep{Papics2014} and KIC  77606790 \citep{Papics2015}.
Detections of period spacing patterns have been presented in a variety of contexts since then \citep{Saio2015, VanReeth2015, Murphy2016, Papics2017, Ouazzani2017, Pedersen2021, Szewczuk2021}.
These observations have prompted significant investigation into their implications for our understanding of stellar structure and evolution.

Internal rotation has been investigated in \gdor stars \citep{VanReeth2016, Ouazzani2017, Ouazzani2020, Saio2018, Saio2021, Li2019b, Li2020b}.
Detailed studies of KIC~10526294 have also been performed by \citet{Moravveji2015} and \citet{Zhang2023} into the nature and shape of convective overshooting.
Another well-studied SPB star, KIC~7760680, was explored in detail \citep{Moravveji2016, Michielsen2021, Bowman2021} in the context of core overshooting, envelope mixing, and the temperature gradient in the core boundary layer.
The stellar properties of the SPB stars HD 50230 \citep{Wu2019}  and KIC 8324482 \citep{Wu2020} were derived by comparing their period spacing patterns to stellar models.
More recently, \citet{Pedersen2021} presented period spacing patterns for a large sample of 26 SPB stars, using them to investigate global mixing properties. They compared the observed period spacing patterns to a large grid of stellar evolution models with different assumptions for internal mixing and found a wide range of mixing levels across the sample.

Obtaining information about the stellar structure from oscillation frequencies in studies of SPB and \gdor stars can be difficult.
The usual approach is to compare to a grid of stellar evolution models, compute the oscillation frequencies at different points during the evolution, and then compare the results with the observed frequencies. This is often termed `forward modelling'. Forward modelling has been applied to SPB and \gdor stars in all of the studies discussed above to constrain the structure.

An alternative approach is to compute a static model that best reproduces the observations by iteratively improving on a starting guess for the structure.
This would be complementary to forward modelling studies based on a grid search of evolutionary models.
Such approaches have been investigated in a variety of contexts.
Helioseismic non-linear inversions have been investigated by \citet{Antia1996} and \citet{Marchenkov2000}.
Inversions based on the structure have been applied to white dwarfs  \citep{Giammichele2017a, Giammichele2017b, Giammichele2018}.
Tests of the forward modelling approach on sdB stars have been performed by \citet{Charpinet2008} and on white dwarfs by \citet{Charpinet2019}.
Works on inversions have also been completed by \citet{Roxburgh2000, Roxburgh2002b, Roxburgh2003a} and \citet{Roxburgh2002a}.
Rotation inversions of KIC 11145123 were performed by \citet{Hatta2019, Hatta2021, Hatta2022}.
\citet{Vanlaer2023} studied the feasibility of structure inversions for gravity-mode pulsators based on the variational principle and found that the non-linear dependences of the oscillation frequencies made it too difficult to apply this method to such stars.
We take a non-linear non-variational approach to the inversion problem and present applications in the context of SPB stars.

In this paper we present a method for the non-linear inversion of the stellar structure of gravity-mode pulsators such as SPB and \gdor stars. We begin by describing our inversion procedure in detail in Sect. \ref{method}. 
We apply our inversion procedure to a model inside the parameter space (Sect. \ref{sec_conv_in}), a model outside the parameter space (Sect. \ref{sec_conv_out}), and an observation (Sect. \ref{sec_conv_obs}). We then discuss the advantages and limitations of our inversion procedure in Sect. \ref{discussion} and conclude in Sect. \ref{conclusion}.

\section{Computational method} \label{method}
Each step of our inversion procedure involves the calculation of stellar structure models and pulsation models, followed by a method to update the stellar structure (Fig. \ref{method_summary}). In this section we discuss the details of these calculations.

\subsection{Stellar structure calculations}
Our method involves the calculation of static stellar structure models based on the total mass and internal abundance profiles of the chemical elements.
To compute these static stellar models, we used r22.05.1 of the \mesa software package \citep{Paxton2011, Paxton2013, Paxton2015, Paxton2018, Paxton2019, Jermyn2023}, and solved the stellar structure equations with the same method as in \citet{Farrell2022}. 
The following is a  brief summary of the steps we took:

\begin{enumerate}
\item We began with a starting model with similar internal abundance profiles to the desired solution. To facilitate this, we had a pre-computed grid of stellar models covering masses from 2 to 8 \msun in steps of 0.5~\msun, central hydrogen abundances from 0.70 to 0.05 in steps of 0.05, and metallicities of Z=0.004, 0.009, 0.14 and 0.020.
\item We modified the model file directly to input the desired chemical abundance profiles.
\item We loaded this model file into \mesa and turned off all time-dependent processes, such as the change in abundances due to nuclear reactions, chemical mixing, and mass loss, to find the static solution to the stellar structure equations.
\end{enumerate}
Numerical tests verify that the solution to the stellar structure equations does not depend on the initial model, as long as \mesa is able to find a solution.
To compute these models, we used moderately high spatial resolution ($\sim$8000 zones) with a particular emphasis on the regions close to the abundance gradient. These regions are the most sensitive to the period spacing patterns of the SPB and \gdor stars and a high resolution allows for greater precision in this region. Further details of the physical inputs to our \mesa models, including the equation of state, opacities, and nuclear reaction rates, are discussed in Appendix \ref{appen:mesa}.

\subsection{Oscillation calculations}
To model the oscillation frequencies, we used the \gyre oscillation code \citep{Townsend2013, Townsend2018}. We computed oscillation models under the adiabatic approximation, which is considered appropriate for g-modes dominant deep in the stellar interior. We used the traditional approximation for rotation, considered to be appropriate for SPB and \gdor stars \citep{Ouazzani2017, VanReeth2018}, varying the rotation rate as a fraction of the surface critical rotation rate. We also assumed rigid rotation.

\begin{figure}[h] \centering
\includegraphics[width=0.95\columnwidth]{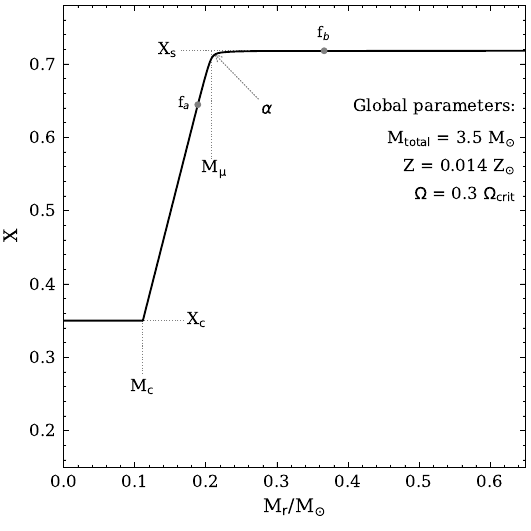}
\caption{Our parameterisation of the hydrogen abundance profile and the three global stellar parameters that we also consider. The parameters in the figure are the mass (M$_{\rm total}$), the metallicity ($Z$), the rotation $\Omega/\Omega_{\rm crit}$, the central and surface hydrogen abundances ($X_c$ and $X_s$), the normalised mass coordinates at the outer boundary of the composition gradient and the convective core ($M_{\mu}$ and $M_c$), and the curvature parameters $\alpha$, $f_a$, and $f_b$. } 
\label{h1_profile}
\end{figure}

\begin{figure}[h] \centering
\includegraphics[width=0.95\columnwidth]{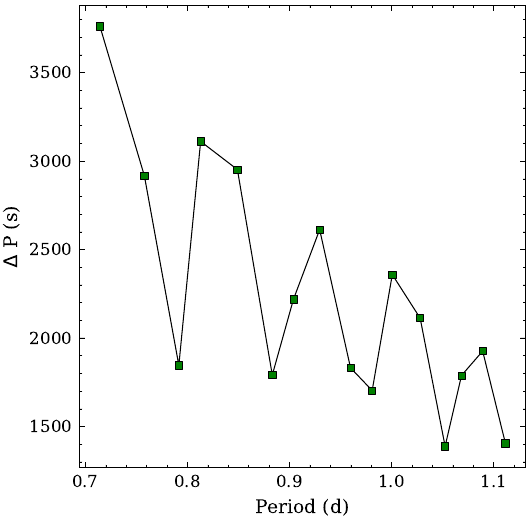}
\caption{Period spacing pattern calculated based on the hydrogen abundance profile in Fig. \ref{h1_profile}.} 
\label{ps_pattern}
\end{figure}

\begin{figure}[h] \centering
\includegraphics[width=0.95\columnwidth]{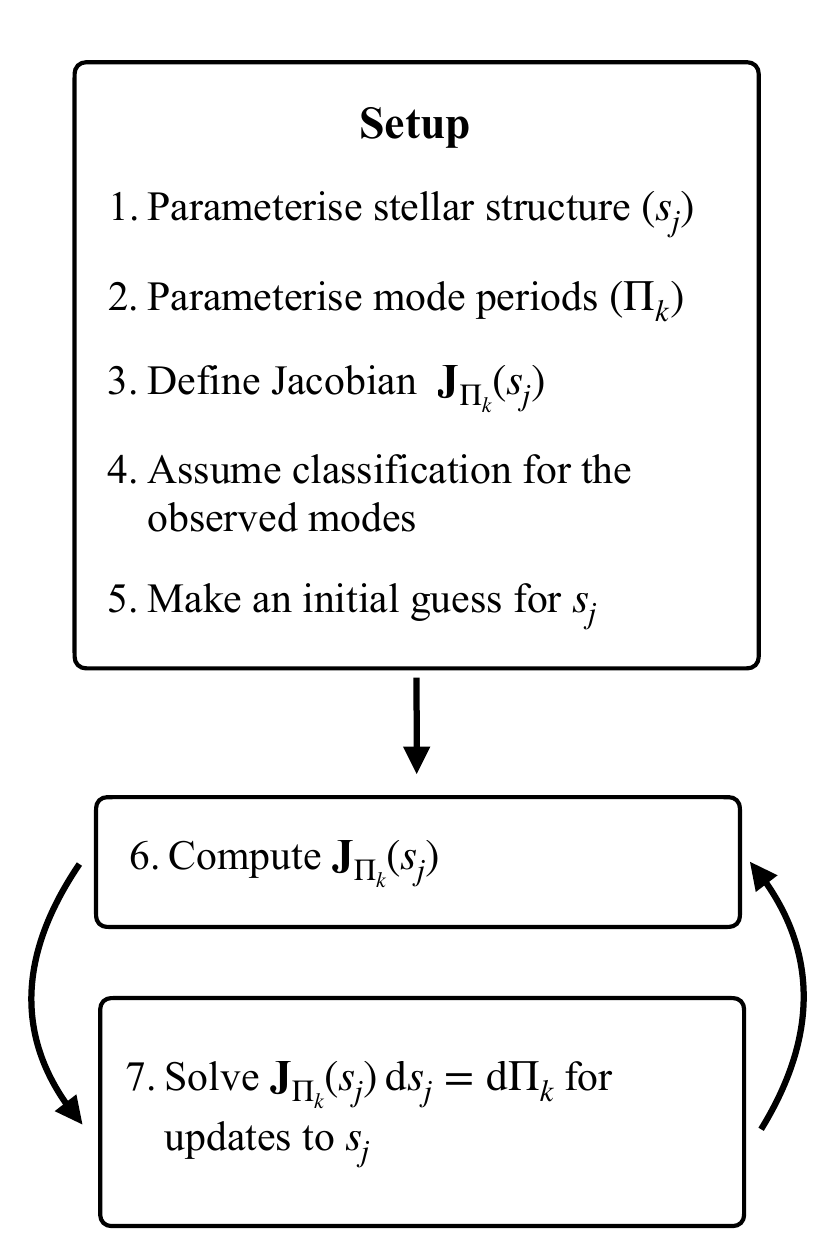}
\caption{Summary of our inversion procedure.} 
\label{method_summary}
\end{figure}

\subsection{Selection of asteroseismic constraints} \label{sec_constraints}

\begin{figure}[h] \centering
\includegraphics[width=0.95\columnwidth]{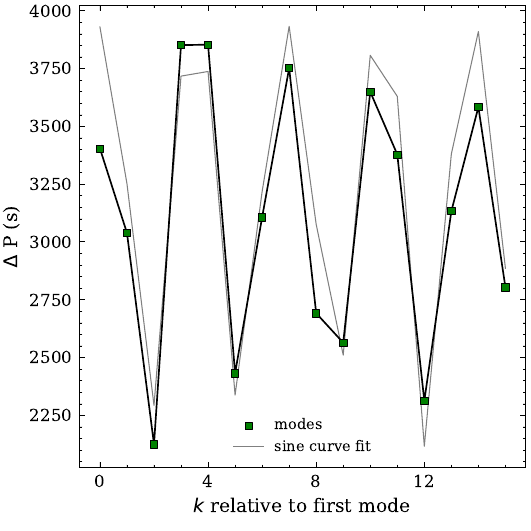}
\caption{Period spacing pattern from Fig. \ref{ps_pattern} modified to exclude the effect of the slope due to rotation.} 
\label{ps_pattern_rot}
\end{figure}

\begin{figure}[h] \centering
\includegraphics[width=0.95\columnwidth]{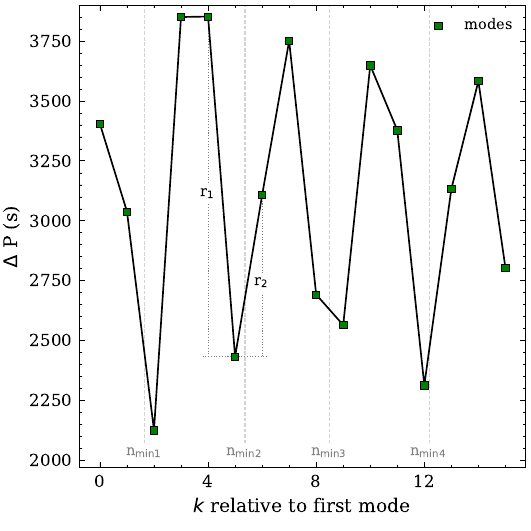}
\caption{Sketch of how we estimate the position of the minimum of each dip in mode number based on the shape of the period spacing pattern.} 
\label{ps_pattern_mins}
\end{figure}

\new{For SPB stars, one typically looks at the mode frequencies in terms of the period spacing patterns \citep{Miglio2008}.
Figure~\ref{ps_pattern} shows the period spacing pattern corresponding to the stellar structure model presented in Fig. \ref{h1_profile}, for prograde dipole modes from k = 14 to 29.}
The mode periods can be parameterised in a variety of ways.
The asymptotic approximation for the mode periods in a rotating star \citep{Bouabid2013} is 

\begin{equation} \centering
P_k = \frac{2\pi^2 (k + \frac{1}{2})}{\sqrt{\lambda_{l, m, \nu(k)}} \int_{r_0}^{r_1}{\frac{|N|}{r}dr}},
\end{equation}

\noindent where $\int_{r_0}^{r_1}{\frac{|N|}{r}dr}$ is the buoyancy radius, $\lambda_{l, m, \nu(k)}$ are the eigenvalues of the Laplace tidal equations and where $\lambda_{l, m, \nu(k)} = l(l+1)$ in the absence of rotation. This expression implies a constant asymptotic period spacing for non-rotating stars.

\citet{Miglio2008} presented an analytical analysis of the impact of a jump in the \bv profile on the period spacing patterns, demonstrating that mode trapping leads to dips in the period spacing patterns with characteristics that related to the \bv profile.
They derived the following analytical expression for $\Delta P$ based on a \bv profile with a step function discontinuity that mimics the \bv profile produced by a partially burned convective core:

\begin{equation} \label{eqn_deltap}
\Delta P \propto \frac{\Pi_{0}}{L} \frac{1 - \alpha^2}{\alpha^2} \cos \biggl( 2 \pi  \frac{\Pi_{0}}{\Pi_{\mu}}k + \pi \frac{\Pi_{\mu}}{\Pi_{0}} \phi  + \frac{\pi}{2} \biggl),
\end{equation}where $\Pi_{\mu}$ is the buoyancy radius at the discontinuity, $\Pi_{0}$ is the total buoyancy radius, $k$ is the radial order, $L=[l(l+1)]^{1/2}$, and $\alpha$ parameterises the jump in \bv frequency such that $\alpha=1$ has no jump and $\alpha \rightarrow 0$ has a sharp discontinuity (see \citealt{Miglio2008} for further details).

The expression for $\Delta P(k)$ in Eq.~(\ref{eqn_deltap}) contains a sinusoidal component with a periodicity in terms of the radial order given by $\Pi_{\mu}/\Pi_{0}$, an amplitude that is proportional to the sharpness of the variation in the \bv profile and that does not depend on $k$. For a ramp function \bv profile, one finds a similar expression for $\Delta P$ with an amplitude that decreases with increasing $k$ \citep{Miglio2008, Hatta2023}.
This demonstrates that the periodicity and amplitude of the dips in the period spacing pattern contain useful information about the position and curvature of the outer edge of the peak of the \bv frequency profile in normalised buoyancy radius, or the outer edge of the hydrogen gradient.

We first considered parameterising the periodicity of the dips in the period spacing pattern by fitting a sine curve, once the slope due to rotation is accounted for.
However, we were unable to find the period of the best-fit sine curve with sufficient precision in a reliable and consistent way.
This is consistent with the well-known fact that best-fit parameters of non-linear functions are known to be strongly dependent on the initial guess in a highly non-linear way.
We tried unsuccessfully with a function based on the absolute value of a sine curve.
We also tried a quadratic fit to the three period spacings closest to the minima, but also found this to be unreliable.
Ultimately, we decided on a more deterministic procedure to calculate the positions and periodicity of the dips in the period spacing patterns.

Our procedure first required us to modify the period spacing pattern due to the slope caused by rotation.
We achieved this by fitting the period spacings with an inverse cubic function as a function of mode number, dividing the period spacings by this fit, and then scaling by a constant factor (see Appendix \ref{rotation_correction} for further discussion).
An example of the output of this procedure is the pattern shown in Fig. \ref{ps_pattern_rot}.
The advantage of this procedure is that it produces approximately the same pattern regardless of the initial rotation.
For the amplitude, we found that simply fitting the sine curve to the resultant pattern is sufficiently reliable and so used that approach.
For the periodicity, we continued by computing the difference of adjacent $\Delta P$, indicated by $r_1$ and $r_2$ in Fig. \ref{ps_pattern_mins}.
We then computed

\begin{equation} \label{eqn_min}
n_{\rm min, 2} = n_{\rm orig} + 0.5 \times \cos\Big(\frac{\pi r_2}{2 r_1}\Big)
,\end{equation}

\noindent where $n_{\rm orig}$ corresponds to the radial mode number of the actual minimum of the period spacing pattern (Fig. \ref{ps_pattern_mins}) and $n_{\rm min, 2}$ corresponds to our defined minimum of the second dip of the period spacing pattern. As desired, this procedure would result in a value of $n_{\rm min, 2} = 5$ if $r_1 = r_2$ and $n_{\rm min, 2} = 5.5$ in the limit that $r_2 = 0$. 
Repeating this procedure for all minima and calculating differences between each minimum gives us the periodicity of the dips in a reliable deterministic way that can be used in our inversion procedure.
Ultimately, we need a constraint with the property that it exhibits a monotonic dependence with respect to a change in the structure parameters and we find that Eq. \ref{eqn_min} exhibits this property (see Appendix \ref{app_constraints_params} for further details).
We note that one may have to be careful when dealing with observations at longer periods, where signal to noise is lower and mode classification is more difficult, which may result in spurious minima appearing in the period spacing pattern that may not be physically linked to normalised buoyancy radius of the \bv discontinuity.

Another possible set of asteroseismic constraints is simply to use the periods directly.
In practice, we find a combination of different constraints to be useful, depending on how far away the model frequencies are from the target frequencies, and on which physical parameters are being constrained.
When close to the target solution, the periods and/or period spacings are sometimes useful.
We find that using just the values of the shortest period, $P_0$, and longest period, $P_{n_P}$, can be used to constrain the mass and rotation, capturing information about the buoyancy radius and the slope. Again, we chose not to parameterise based on the slope of the period spacing pattern directly due to non-linearities.

\subsection{Parameterisation of the stellar structure}
Figure \ref{h1_profile} presents our parameterisation of the hydrogen abundance profile for a model of a  3.5 \msun core hydrogen burning star, representative of SPB stars. We selected this parameterisation to try to capture the important structural components that occur in stellar evolution models of such stars. The parameterisation consists of three straight lines, joined by a sharp angle at the core boundary and with a curve at the outer edge of the abundance gradient region. The hydrogen abundance profile in the core is a straight line as it is convective. The intermediate zone between the core and envelope is also a straight line. This region is formed over a nuclear-burning timescale by the receding convective core. The envelope is flat, as is typically expected from stellar evolution models. However, the boundaries of the curved connection between the gradient and the flat envelope (see Appendix \ref{append_curve}) can allow for profiles with extra envelope mixing produced by additional physical processes \citep[e.g.][]{Aerts2019}.

Our parameterisation contains four parameters related to the global shape of the hydrogen abundance profile. These are (i) the central hydrogen abundance $X_c$; (ii) the surface hydrogen abundance $X_s$; (iii) the normalised mass coordinate at the outer edge of hydrogen gradient $M_{\mu}$; and (iv) slope of the hydrogen gradient $H_{\rm slope}$, defined as 

\begin{equation}
H_{\rm slope} = \frac{X_s - X_c}{M_{\mu} - M_c}
,\end{equation}

\noindent where $M_c$ is the normalised mass coordinate of the core as indicated in Fig \ref{h1_profile}. 
In addition, our parameterisation contains three global parameters. These are (i) the total stellar mass, $M$; (ii) the rotation as a fraction of the critical rotation, $\Omega/\Omega_{\rm crit}$; and (iii) the metallicity, $Z$. The metallicity is potentially difficult to parameterise. In this work, we simply constructed pre-computed grids at discrete metallicities (0.003, 0.009, 0.014, and 0.020) and then scaled the metal abundance profiles as follows. In the convective core and the envelope, we assumed a constant abundance of each metal with a value equal to that of the closest pre-computed model (i.e. closest in $M$ and $X_c$) to account for the evolution of the core abundances of $^{12}$C, $^{14}$N, and others. In the transition region consisting of the hydrogen gradient, we took the metal abundance profiles of the closest pre-computed model and scale as a function of the normalised mass in this region.

Finally, to mimic hydrogen profiles produced by internal mixing processes in stellar evolution models (e.g. rotation), we included a curved parameterisation of the connection between the hydrogen gradient and the flat outer envelope that is defined by three parameters. The parameters $f_a$ and $f_b$ define the fractional positions of the hydrogen gradient and the envelope at which the curve begins and ends. Based on general comparisons with stellar evolution models, we often chose values of $f_a = 0.8$ and $f_b = 0.2$ and left these constant for simplicity. The curvature is defined by $\alpha$, where $\alpha = 0$ results in a sharp connection between the two lines (i.e. no curve at all) and $\alpha = 1$ results in a straight line between $f_a$ and $f_b$. Stellar evolution models with extra diffusive mixing in the envelope typically correspond to $\alpha \sim 0.05 - 0.10$. The exact equation used to define the curve is detailed in Appendix \ref{append_curve}, along with examples of the parameterisation. For now, we ignored deviations from spherical symmetry in the hydrogen abundance profile. We chose not include a curvature at the connection between core profile and the hydrogen gradient in our parameterisation as our tests showed that this was less important for the overall structure that the curve at the upper edge of the gradient, and also because stellar evolution models typically do not produce such a feature.

To avoid the parameterisation generating unphysical stellar models, we set minimum and maximum values for each of these parameters. These include constraints that $0.01 < X_c < 0.72$, $0.05 < M_{\mu} < 0.50$, and $2~\msun < M < 8~\msun$. These constraints are intended to be relatively general, but can be adjusted depending on the context of the star being studied. We also required that the hydrogen abundance increases moving outwards, $X_c \lesssim X_s$. In our inversion procedure, we typically allowed $M$, $\Omega/\Omega_{\rm crit}$, $X_c$, $M_{\mu}$, and $\alpha$ to vary, keeping $f_a$, $f_b$, $Z$, $H_{\rm slope}$, and $X_s$ fixed due to degeneracies. We can explore the impact of, for example, the metallicity by changing the fixed value of $Z$ and repeating the inversion procedure.

An alternative approach would have been to parameterise the stellar structure in terms of the \bv frequency profile instead of the hydrogen abundance profiles.
The hydrogen abundance profile has a more direct link with the stellar structure, while the \bv frequency profile has a direct link with the oscillation frequencies.
The hydrogen abundance profile has the advantage that it is easier to reject unphysical models and that the results are immediately interpretable in terms of the physics of stellar evolution (i.e. mixing processes). The \bv profile may require a more intricate parameterisation, be more prone to unphysical solutions and may require more parameters to capture the full shape of the profile. The \bv frequency profile is also independent of other physical assumptions regarding the equation of state, opacities, nuclear reactions etc.
In this work, we chose to parameterise based on the hydrogen abundance profile for the application to SPB and \gdor stars. We plan to explore parameterisations of the \bv frequency profile in future work.
We emphasise that our method may be improved upon with specific parameterisations for different contexts.

\subsection{Overview of the inversion procedure}
Figure \ref{method_summary} summarises our inversion procedure for finding the stellar structure model that best matches the mode periods. In summary, the procedure requires the following steps:

\begin{enumerate}
\item Parameterise the stellar structure based on the internal hydrogen abundance profile with parameters $s_j$.

\item Parameterise the observed mode periods or frequencies (e.g. the periodicity and amplitude of the dips that appear in the period spacing pattern) with parameters $\Pi_k$.

\item Define the Jacobian $\mathbf{J}_{\Pi_k}(s_j)$.

\item Assume an identification for the observed modes.

\item Make an initial guess for the structure parameters $s_j$ and compute the stellar structure and oscillation frequencies for this initial guess.

\item Construct a Jacobian matrix of partial derivatives $\mathbf{J}_{\Pi_k}(s_j)$ by modifying each structure parameter $s_j$ one by one and re-computing the stellar structure and oscillation frequencies.

\item Use the Jacobian to solve $\mathbf{J}_{\Pi_k}(s_j) \, \mathrm{d} s_j = \mathrm{d} \Pi_k$ for the corrections $\mathrm{d} s_j$ that should be applied to $s_j$.

\item Update the trial solution and repeat steps 6 and 7 until the corrections to the structure parameters $\mathrm{d} s_j$ fall below some desired threshold.

\end{enumerate}

\subsection{Dependence of constraints on the stellar structure}

Given a series of structure parameters $s_j$ and a series of asteroseismic constraints $\Pi_k$, we would not necessarily expect all the structure parameters to significantly affect all the constraints.
For this reason, we chose to set some of the entries in the Jacobian to 0.
This decision can be made both to aid numerical convergence and based on physical arguments.
As an example of a physical argument, if we changed the curvature of the gradient, parameterised by $\alpha$, we would expect this to change the amplitude of the dips in the period spacing significantly, but not the periodicity of the dips \citep{Miglio2008}.
Numerically, we found that small changes in the other constraints can lead to large gradients in the Jacobian matrix, which can cause difficulties with numerical convergence in the inversion process.
We find that setting some of these elements to 0 was a key step in helping the inversion method converge towards a solution.
Divergence in this context primarily manifested by parameters consistently hitting the boundaries of the parameter space.

\subsection{Mode identification}
Our method requires the mode identification as an input.
For the mode classification of SPB stars, the angular degree ($l$) and azimuthal order ($m$) are often inferred by manually building a pattern from a list of frequencies to satisfy the approximate period differences expected, and then by fitting the period spacing pattern following, for example, \citet{VanReeth2016, VanReeth2018}.
However, there is always an intrinsic degeneracy due to the unknown radial order of the modes.
Our approach in this case is to first make a guess for the identification of the modes based on comparisons with stellar models and perform the inversion process, and then to repeat the inversion process with different assumptions for the mode identification (Fig. \ref{degeneracy_example_mode} shows an example of the degeneracy that one can encounter due to the unknown radial order of the mode).
This method is versatile and has the advantage of explicitly exploring possible degeneracies in the mode identification.
However, it does mean that one may have to perform the inversion process several times.
Dealing with the unknown mode classification would need to be done on a case-by-case basis, depending on the number of modes available, the evolutionary regime and non-seismic constraints.
We intend to explore degeneracies related to the unknown radial order in SPB stars in a future publication.

\subsection{Initial guess for the structure parameters}
As always, these kinds of inversion procedures depend on having a reasonably good starting model.
The inversion procedure also requires an initial guess for the structure parameters, $s_j$.
A reasonable first guess can be achieved by comparing with a pre-computed grid of models.
A better initial guess helps with the inversion process; however, our inversion procedure often converges from surprisingly far away in the parameter space.
For instance, on artificial targets within the parameter space, the mass could be changed by $\sim$1~\msun, $X_c$ by $\sim$0.20, $M_{\rm \mu}$ by $\sim$0.1, and $\Omega/\Omega_{\rm crit}$ by $\sim$0.3 and the procedure would still converge to the correct solution.
However, for models outside the parameter space or with observations, the initial guess needed to be closer to the final solution in order to converge. In this case, the mass could be changed by approximately $\sim$0.3~\msun, $X_c$ by $\sim$0.08, $M_{\rm \mu}$ by $\sim$0.05, and $\Omega/\Omega_{\rm crit}$ by $\sim$0.2.

\subsection{Computing the Jacobian and solving for updates to $s_j$}
During convergence, each iteration begins with a trial solution and is updated using the Jacobian matrix.
Computing the Jacobian is relatively straightforward. 
The structure parameters, $s_j$, are modified one at a time, keeping all the other structure parameters constant, and the stellar structure and corresponding oscillations are computed.
For the first iteration, we chose a small change for each $s_j$, for instance a change of 0.02 in $X_c$ and 0.02 in the mass, $M$. For subsequent iterations, we used the Jacobian from the previous iteration to compute the expected magnitude of the change in each $s_j$. This allows the inversion procedure to take smaller steps in $s_j$ in the calculation of the Jacobian when the model is closer to the solution.
For each $s_j$, the changes to the stellar structure were computed with \mesa, the corresponding oscillation frequencies were computed with \gyre, and the asteroseismic constraints, $\Pi_k$, were computed.
They provided the columns of the Jacobian as given by Eq~(\ref{jacobian}):

\renewcommand{\arraystretch}{2}

{\Large

\begin{equation} \label{jacobian}
\begin{bmatrix}
\frac{\partial A_{1}}{\partial S_{1}} & \frac{\partial A_{1}}{\partial S_{2}} & \cdots & \frac{\partial A_{1}}{\partial S_{n_{\mathrm{s}}}}\\
\frac{\partial A_{2}}{\partial S_{1}} & \frac{\partial A_{2}}{\partial S_{2}} & \cdots & \frac{\partial A_{2}}{\partial S_{n_{\mathrm{s}}}}\\
\vdots & \vdots & \ddots & \vdots \\
\frac{\partial A_{n_{\mathrm{a}}}}{\partial S_{1}} & \frac{\partial A_{n_{\mathrm{a}}}}{\partial S_{2}} & \cdots & \frac{\partial A_{n_{\mathrm{a}}}}{\partial S_{n_{\mathrm{s}}}}\\
\end{bmatrix}
\begin{bmatrix}
dS_{1} \\ dS_{2}  \\ \dots \\ dS_{n_\mathrm{s}} \\ 
\end{bmatrix}
=
\begin{bmatrix}
dA_{1} \\ dA_{2}  \\ \dots \\ dA_{n_\mathrm{a}} \\ 
\end{bmatrix}
,\end{equation}

}

\noindent or in a more concise form,
\begin{equation} \label{jacobian_condensed}
\mathbf{J}_{\Pi_k}(s_j) \, \mathrm{d} s_j = \mathrm{d} \Pi_k
\end{equation}Equation~\ref{jacobian_condensed} is solved using the least squares method; this provides the required updates to the structure parameters, $\mathrm{d} s_j$, by minimising 
$|\mathbf{J}_{\Pi_k}(s_j) \, \mathrm{d} s_j - \mathrm{d} \Pi_k|$.
The trial solution to the stellar structure equations is then updated based on this change in $s_j$ and the Jacobian is recomputed.
These steps are repeated until the structure parameters converge towards a desired tolerance.
The procedure allows us to begin with one set of $\Pi_k$ and Jacobian structure, and then to change to a different set of $\Pi_k$ once the trial solution becomes closer to the target solution. We find this useful, as the set of $\Pi_k$ that aid convergence when far away from the solution are sometimes different to the set of $\Pi_k$ that aid convergence when close to the solution.

\section{Examples of applying of our inversion technique}

In this section we test our inversion technique on a static model parameterised with the same parameters that are being varied by the inversion procedure (Sect. \ref{sec_conv_in}), a snapshot taken from a stellar evolution model (Sect. \ref{sec_conv_out}), and an observation (Sect. \ref{sec_conv_obs}). For the two tests with artificial data from models, we considered examples with similar data quality as the best SPB stars from \textit{Kepler} and also assumed the correct radial order identification. For the observation, the radial mode orders are not known, so we made an assumption for the mode identification. The consequences of exploring degeneracy related to different mode assumptions are complex (see e.g. Fig. \ref{degeneracy_example_mode}) and we leave that to a further detailed study of this star.

\subsection{Convergence examples of models inside the parameter space} \label{sec_conv_in}

We now provide a representative example of how our inversion method works for artificial data computed from a static model parameterised with the same parameters that are being varied by the inversion procedure.
This is a test case in which our method should ideally perfectly recover the solution.

The target period spacing pattern was computed from a parameterised stellar structure model with a mass of 3.5~\msun star, $\Omega/\Omega_{\rm crit} = 0.30$, $X_c = 0.35$, and $\alpha = 0.05$, For simplicity, we left the other parameters (those presented in Fig. \ref{h1_profile}) constant. We began with a starting guess of a mass of 3.2~\msun star, $\Omega/\Omega_{\rm crit} = 0.10$, $X_c = 0.30$, and $\alpha = 0.03$. We note that for the purposes of this test, we assumed the correct identification of the radial order. We used a Jacobian based on the amplitude of the dips of the period spacing pattern, the first and last periods and the gaps between the minima in the period spacing pattern.

\begin{figure*}[h] \centering
\includegraphics[width=0.99\columnwidth]{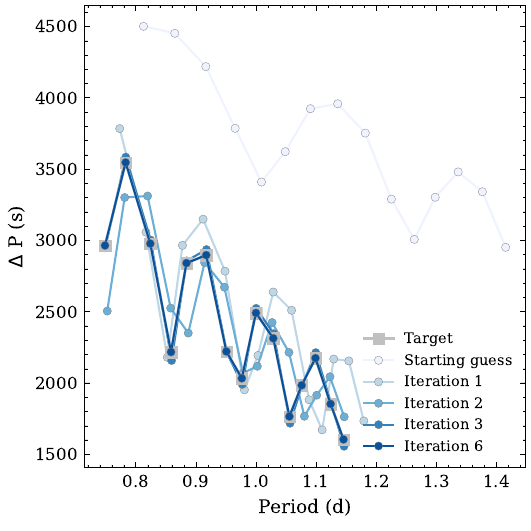}
\includegraphics[width=0.99\columnwidth]{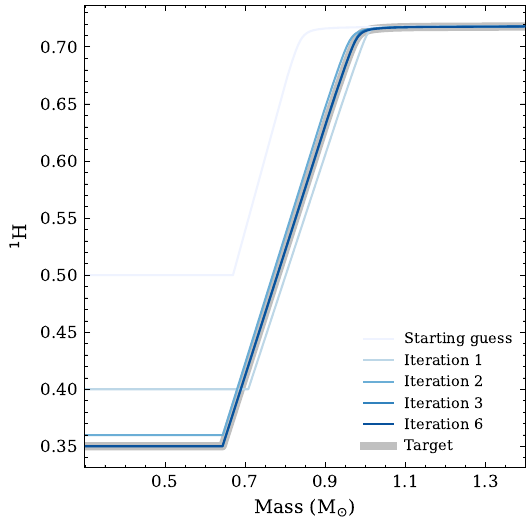}
\caption{{\it Left panel:} Example of the convergence of our inversion procedure for a model inside the parameter space.
{\it Right panel:} Convergence of the hydrogen abundance profile for the same model. Only the region close to the hydrogen region is shown, for clarity.} 
\label{conv_inside_param_space}
\end{figure*}

Table \ref{table_conv_in} shows the values of each structure parameter for each iteration from the starting guess to the final iteration, along with the actual values of the target solution. Figure \ref{conv_inside_param_space} shows the period spacing pattern and the hydrogen abundance profile for each iteration, compared to the target solution.
Figure \ref{conv_inside_param_space_paramplot} shows the convergence within the parameter space for the values in Table \ref{table_conv_in}. 

After five iterations, the inversion procedure recovers values very close to the target solution. The method converges to the exact solution after three further iterations. This is encouraging, because it suggests that if our parameterisation of the stellar structure is good enough and the mode identification is correct, we should be able to converge towards the correct solution. Even though there are only four parameters being varied, it is worth recognising that it is very difficult to achieve the outcome of this experiment by hand as there are so many interaction effects between each structure parameter that are extremely difficult to take into account by hand, but are relatively easily taken into account by the Jacobian. Once the number of parameters is increased to six or more, it becomes substantially more difficult to simultaneously converge to the solution globally. This is because intrinsic degeneracies between the stellar structure and the period spacing patterns are introduced. These degeneracies are recognisable by the production of multiple sets of period spacing patterns produced by different combinations of $s_j$. Most of these degeneracies are related to the fact that there are multiple hydrogen abundance profiles that produce similar \bv frequency profiles and similar buoyancy radii and, therefore, similar period spacing patterns. An example of such a degeneracy is presented in Appendix \ref{degeneracy_example}.

\renewcommand{\arraystretch}{1.5}

\begin{table}[h!]
\centering
\caption{Parameters at each iteration for the test model inside the parameter space.}
{\footnotesize
\begin{tabular}{l rrrr}
 {\bf Iteration \#} &{\bf Mass (M$\bm{_{\odot}}$)} & $\bm{\Omega/\Omega_{\rm crit}}$ & $\bm{X_c}$ & $\bm{\alpha}$ \\  \midrule
Starting Guess       & 3.0000       & 0.1000       & 0.5000       & 0.0800       \\ \midrule
1                    & 3.5930       & 0.2699       & 0.4000       & 0.0309       \\ \midrule
2                    & 3.4424       & 0.2914       & 0.3598       & 0.0556       \\ \midrule
3                    & 3.5022       & 0.2998       & 0.3502       & 0.0427       \\ \midrule
4                    & 3.5020       & 0.3004       & 0.3500       & 0.0503       \\ \midrule
5                    & 3.5000       & 0.3000       & 0.3500       & 0.0500       \\ \midrule
6                    & 3.4999       & 0.3000       & 0.3500       & 0.0500       \\ \midrule
7                    & 3.4998       & 0.3000       & 0.3500       & 0.0500       \\ \midrule
8                    & 3.5000       & 0.3000       & 0.3500       & 0.0500       \\ \midrule
{\it Actual values}  & {\it 3.5000      } & {\it 0.3000      } & {\it 0.3500      } & {\it 0.0500      } \\ \midrule
\label{table_conv_in}
\end{tabular}}
\end{table}

\subsection{Convergence examples of models outside the parameter space} \label{sec_conv_out}

\begin{figure*}[h] \centering
\includegraphics[width=0.99\columnwidth]{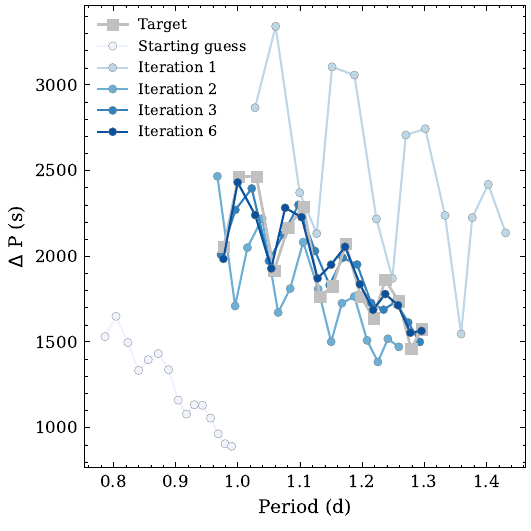}
\includegraphics[width=0.99\columnwidth]{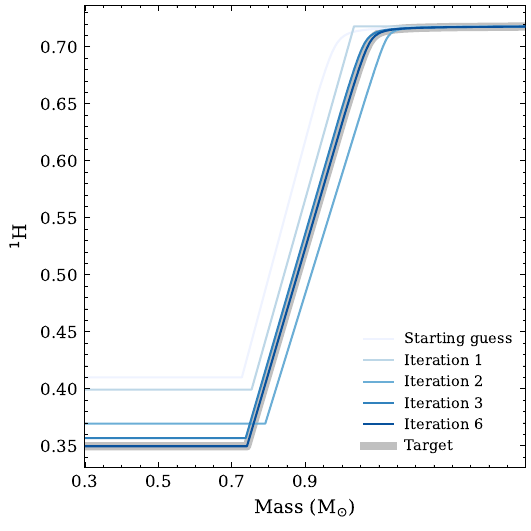}
\caption{{\it Left panel:} Example of the convergence of our inversion procedure for a model outside the parameter space. The model is recovered after six iterations.
{\it Right panel:} Convergence of the hydrogen abundance profile for the same model. Only the region close to the hydrogen region is shown, for clarity. } 
\label{conv_outside_param_space}
\end{figure*}

We now present the results of an experiment using a model outside the parameter space (see Fig.~\ref{conv_outside_param_space} and Table~\ref{table_conv_out}). We extracted a snapshot from a stellar evolution model of a 3.5~\msun star halfway through core hydrogen burning and chose a value for the rotation of $\Omega/\Omega_{\rm crit} = 0.30$. In the stellar evolution model, we generated a curvature in the hydrogen abundance gradient by adding a constant diffusion coefficient in the envelope, as in \citet{Pedersen2021}. We began with a starting guess of a mass of 3.2~\msun, $\Omega/\Omega_{\rm crit} = 0.40$, $X_c= 0.41$, $\alpha = 0.10$ and kept the other parameters constant. The path through the parameter space is plotted in Fig. \ref{conv_outside_param_space_paramplot}, with the values recorded in Table \ref{table_conv_out}. Figure \ref{conv_outside_param_space} shows the period spacing pattern for selected iterations. In this example, we used a Jacobian that depended on the amplitude, first and last periods and the periodicity of the dips in the period spacing pattern for the first four iterations and then included a dependence on the actual positions of the dips for the last two iterations.

We find that our inversion procedure recovers the values of the mass, rotation and central hydrogen abundance reasonably well. As the target model is a snapshot from a stellar evolution model, there is no exact correspondence with $\alpha$. However, a quick comparison of the hydrogen abundance profiles by eye shows the value to be reasonable (Fig. \ref{conv_outside_param_space}).

\begin{table}[h!]
\centering
\caption{Parameters at each iteration for the test model outside the parameter space.}
{\footnotesize
\begin{tabular}{l rrrr}
 {\bf Iteration \#} &{\bf Mass (M$\bm{_{\odot}}$)} & $\bm{\Omega/\Omega_{\rm crit}}$ & $\bm{X_c}$ & $\bm{\alpha}$ \\  \midrule
Starting Guess       & 3.2000       & 0.4000       & 0.4100       & 0.1000       \\ \midrule
1                    & 3.3531       & 0.2185       & 0.3993       & 0.0000       \\ \midrule
2                    & 3.6491       & 0.3183       & 0.3694       & 0.0525       \\ \midrule
3                    & 3.4559       & 0.2982       & 0.3567       & 0.0746       \\ \midrule
4                    & 3.5990       & 0.3243       & 0.3251       & 0.0639       \\ \midrule
5                    & 3.5168       & 0.3043       & 0.3499       & 0.0605       \\ \midrule
6                    & 3.4989       & 0.3036       & 0.3495       & 0.0722       \\ \midrule
7                    & 3.5292       & 0.3052       & 0.3495       & 0.0705       \\ \midrule
8                    & 3.5296       & 0.3051       & 0.3496       & 0.0714       \\ \midrule
{\it Actual values}  & {\it 3.5000      } & {\it 0.3000      } & {\it 0.3500      } & {\it -} \\ \midrule
\label{table_conv_out}
\end{tabular}}
\end{table}

\subsection{An example with an observation with Kepler data} \label{sec_conv_obs}
\begin{figure*}[h] \centering
\includegraphics[width=0.99\columnwidth]{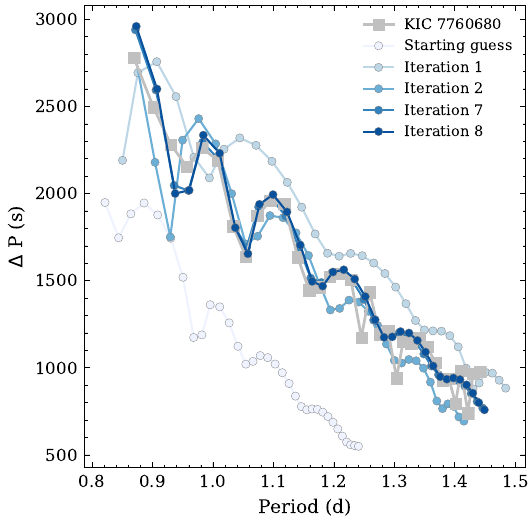}
\includegraphics[width=0.99\columnwidth]{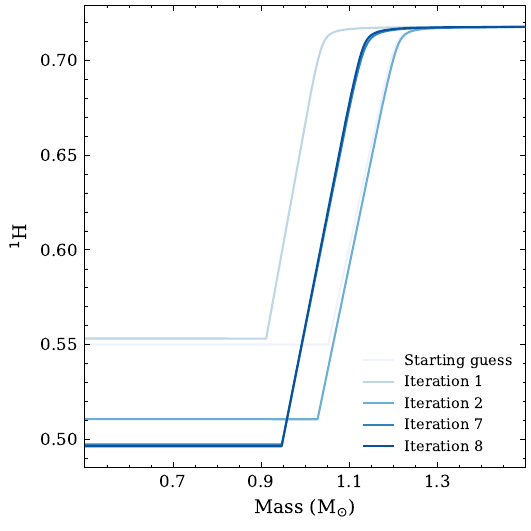}
\caption{{\it Left Panel}: Example of the convergence of the period spacing pattern when applying our inversion technique to the observed modes of the SPB star KIC 7760680 \citep{Papics2015}. {\it Right Panel}: Hydrogen abundance profiles at each iteration.} 
\label{conv_obs}
\end{figure*}

Figure \ref{conv_obs} presents the result of a test of our inversion procedure on the well-studied SPB star KIC~7760680. We emphasise that this test is not meant to be a thorough characterisation of the star, which can be found elsewhere in the literature \citep{Moravveji2016, Papics2015, Michielsen2021, Bowman2021, Pedersen2021}. A detailed characterisation of the star using our technique, taking degeneracies into account, is beyond the scope of this work. However, we can still provide a proof-of-concept that our method works in less ideal conditions than comparing to a stellar model. For the purposes of this test, we allowed five structural parameters to be varied -- (i) the mass, (ii) the rotation, (iii) $\alpha$, (iv) $X_{\rm c}$, and (v) $M_{\rm \mu}$ -- and held the other stellar parameters fixed. In the Jacobian, the rotation, mass and $X_c$ depend on the initial and final periods and the positions of the dips, $\alpha$ depends on the amplitude and $M_{\rm \mu}$ depends on the values of the periods and the positions of the dips.
To generate a starting guess, we took a model from our pre-computed grids that visually approximately matches the slope and period range of the observed modes. Our starting model has $M = 4.0 \msun$, $\Omega/\Omega_{\rm crit} = 0.3$, $\alpha = 0.03$, $X_c = 0.55$, $M_{\mu} = 0.30$, $H_{\rm slope} = 4.6$, $X_s = 0.718,$ and $Z = 0.140$. The $H_{\rm slope}$, $X_s$, and $Z$ are fixed, and the other values for $s_j$ are allowed to vary. We obtain the same solution with different starting models (e.g. combinations of $M = 3.5 \msun$, $\Omega/\Omega_{\rm crit} = 0.2$, $\alpha = 0$ etc.)
The period spacing pattern for the starting guess and three iterations of the inversion are compared to the observed period spacing pattern \citep{Papics2015} in Fig. \ref{conv_obs}. For now, we did not take observational errors into account.
The inversion procedure significantly improves upon the starting guess from a visual comparison of the position and shape of the dips in the period spacing patterns. The corresponding changes to the hydrogen abundance profile are shown in the right panel.

The final values that we obtain for $s_j$ that we allowed to vary are: $M = 3.75 \msun$, $\Omega/\Omega_{\rm crit} = 0.26$, $\alpha = 0.028$, $X_c = 0.496,$ and $M_{\mu} = 0.302$. While the test we present here is not meant to be a proper characterisation of KIC 7760680, it is worth noting that the values for the mass, rotation rate and $X_c$ that we obtain are approximately consistent with previous results \citep{Papics2015, Moravveji2016, Michielsen2021, Pedersen2021}. Our best-fit model hints at some mixing process that is smoothing the gradient, also consistent with previous work.
We emphasise here that this is just a preliminary analysis and a fully study, including characterising degeneracy, will be completed in the future.

\section{Advantages and limitations of our inversion procedure} \label{discussion}

\subsection{Advantages}

Our inversion approach appears to work reasonably well, both on artificial targets and on observations. The method can perfectly recover artificial targets that are inside the parameter space when varying up to five structure parameters. This method also provides a systematic way to characterise the degeneracies involved in the inversion problem. These degeneracies can be due to either the identification of the modes or the intrinsic degeneracies between the stellar structure and the period spacing pattern. In principle, this method could be extended to other types of stars other than gravity-mode pulsators. 
It also appears possible to generalise the approach to other types of stars.
Such a generalisation would likely first require a determination of an optimal parametrisation of the structure based on either the hydrogen or helium abundance profile, or even the \bv frequency profile. Moreover, the method would also need to be adapted to the intricacies of the oscillation spectrum of the target as well as potential surface effects in the case of solar-like oscillations. Some works have been carried out in that respect \citep{Roxburgh2015}, but this would require further investigation.

\new{Compared to the forward modelling approach, a non-linear inversion should allow us to more directly isolate the exact physical information that is available in the period spacing patterns, as well as the limitations due to degeneracy. This is because the output of the inversion procedure identifies the layers in the star that are probed by the period spacings.
The non-linear inversion approach also generally requires the computation of fewer stellar models, although this depends on how targeted the forward modelling grid is. For instance, to setup our inversion method we initially computed 52 stellar evolution models covering our desired mass and metallicity range, and each iteration of the inversion procedure requires the computation of four to five stellar structure models and period spacing patterns. 
Forward modelling has the advantage that the results can be directly connected to time-dependent physical processes, such as convection, and also that they can be immediately placed in an evolutionary context.}

\subsection{Limitations}
There are several possible limitations that need to be kept in mind when applying our method to observations. First, the procedure requires a mode identification as an input. In many cases, this can be difficult or impossible to obtain directly from observations (e.g. the radial order in SPB stars). To overcome this, our strategy is to simply repeat the inversion procedure with multiple different assumptions for the mode identification. This has the added benefit of systematically characterising the degeneracy that may exist due to the mode identification (e.g. Fig. \ref{degeneracy_example_mode}).
Second, the procedure requires an initial guess to begin the inversion procedure. Such an initial guess could be obtained from a grid of stellar evolution models or from a grid of pre-computed static models. To ensure the robustness of the result to the initial guess, we also plan to perform the inversion procedure many times with a range of different initial guesses.
Third, intrinsic degeneracies between the stellar structure and the period spacing pattern will exist (e.g. Fig. \ref{degeneracy_example}). These degeneracies may cause difficulties with convergence. A systematic approach to dealing with such degeneracies is to perform an inversion with fixed values of certain $s_j$ and then change these values and repeat the inversion.
Finally, it is not obvious how exactly to define a good agreement between a model and observations, due to the lack of detailed knowledge of the possible physics that may be at play in the forward problem of computing the frequencies from the stellar structure. New 2D stellar evolution models will help in this regard \citep[e.g.][]{Mombarg2023}.

\section{Conclusions} \label{conclusion}

In this paper we presented a novel non-linear non-variational inversion technique for inferring the internal structure of gravity-mode pulsators such as SPB and \gdor stars. Our approach is based on static stellar models \citep[e.g.][]{Farrell2022} combined with an iterative correction of a parametrised stellar structure, chosen here to be the internal hydrogen profile as a function of the fractional mass coordinate. We have tested our technique on both artificial targets and one \textit{Kepler} observation, KIC 7760680, providing a proof of concept for the technique and for non-linear non-variational inversions that are complementary to stellar evolution models. Our main conclusions are as follows:

\begin{enumerate}
\item A relatively simple parameterisation of a hydrogen abundance profile is sufficient to converge on a solution in a limited number of iterations.
\item Our procedure succeeds in recovering the exact solution when the model is inside the parameter space.
\item We are able to reproduce the period spacing patterns of both artificial data and observations.\item One potential limitation is the degeneracy of solutions based on a given set of observed modes. This means that our method would be restricted to the best possible candidates. On the other hand, our method can be used to systematically investigate degeneracies between the stellar structure and the mode periods.
\end{enumerate}

We plan to use this method in the future to conduct detailed studies of individual SPB stars observed with the \textit{Kepler} mission.
This method can also be applied to other g-mode pulsators, for example \gdor stars and white dwarfs. These non-variational methods are also potentially the natural approach for dealing with structural inversions of other F-type stars, red giants, and core helium burning stars.
As such, our method offers a complementary approach to extensive evolutionary modelling \citep[e.g.][]{Pedersen2021} and shows great potential to constrain the transport processes acting at the border of the convective core of SPB stars, when applied to the best \textit{Kepler} targets.  Such approaches are greatly needed to improve classical 1D stellar evolution models, and reveal their limitations. Conceptually, our method is similar in philosophy to existing non-linear inversion techniques \citep{Giammichele2017a, Antia1996, Marchenkov2000}, and a generalisation to other asteroseismic targets can be envisioned, provided that the formalism and structure parametrisation are adequately adapted. Natural potential targets are F-type solar-like oscillators \citep{Betrisey2023}, \gdor stars \citep{VanReeth2016, Takata2020b}, or sub-giants \citep{Deheuvels2014, Deheuvels2020b} and red giants \citep{DiMauro2018, Vrard2022} that exhibit mixed oscillation modes that have a strong intrinsic non-linear behaviour.

\begin{acknowledgements}
EF and GM have received funding from SNF grant No 200020\_212124.
GM has received funding from the European Research Council (ERC) under the European Union’s Horizon 2020 research and innovation programme (grant agreement No 833925, project STAREX).
 GB has received funding from the SNF AMBIZIONE grant No 185805 (Seismic inversions and modelling of transport processes in stars).
DMB gratefully acknowledges the Engineering and Physical Sciences Research Council (EPSRC) of UK Research and Innovation (UKRI) in the form of a Frontier Research grant under the UK government’s ERC Horizon Europe funding guarantee (SYMPHONY; grant number [EP/Y031059/1]), and a Royal Society University Research Fellowship (grant number: URF/R1/231631). 
EF thanks May Gade Pedersen for sharing observational data and helpful discussions.
\end{acknowledgements}

\appendix 

\section{\mesa} \label{appen:mesa}

The \mesa equation of state is a blend of the OPAL \citep{Rogers2002}, SCVH
\citep{Saumon1995}, FreeEOS \citep{Irwin2004}, HELM \citep{Timmes2000},
and PC \citep{Potekhin2010} equations of state.
Radiative opacities are primarily from OPAL \citep{Iglesias1993,
Iglesias1996}, with low-temperature data from \citet{Ferguson2005}.
Electron conduction opacities are from \citet{Cassisi2007}.
Nuclear reaction rates are from JINA REACLIB \citep{Cyburt2010} with
additional tabulated weak reaction rates from \citet{Fuller1985}, \citet{Oda1994},
and \citet{Langanke2000}.
 Screening is included via the prescription of \citet{Chugunov2007}.
Thermal neutrino loss rates are from \citet{Itoh1996}.
We adopted a mixing length parameter of $\alpha_{\rm MLT} = 1.82$ and used the \texttt{pp\_and\_cno\_extras.net} nuclear reaction network.
The \mesa input files are available as a ZENODO respository: \href{https://zenodo.org/records/10976544}{doi.org/10.5281/zenodo.10976544}.

\section{Equation for the parameterisation of a curved region of an abundance profile} \label{append_curve}

\begin{figure}[h] \centering
\includegraphics[width=0.95\columnwidth]{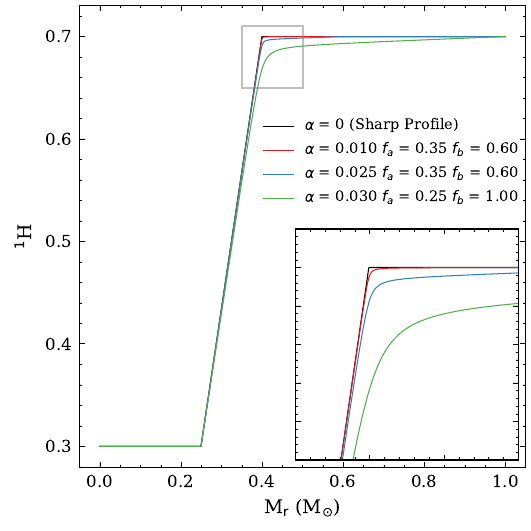}
\caption{Examples of the parameterisation of the curved region.} 
\label{fig_curved_alpha3}
\end{figure}

We used the following equation to parameterise the smooth curve between the hydrogen gradient and the envelope:
\begin{equation} \label{eqn_alpha}
^{1}{\rm H} = \frac{\alpha}{M + \alpha} - \frac{M  \alpha}{1 + \alpha}
,,\end{equation}

\noindent where $M$ is the normalised mass coordinate in the curved region, bounded by $f_a$ and $f_b$. We then applied a rotation of the profile produced by Eq.~(\ref{eqn_alpha}) based on the original angle between the hydrogen gradient and the envelope. This produces the curve shown in Fig. \ref{fig_curved_alpha3}.

\section{Correction for rotation} \label{rotation_correction}
\begin{figure}[h] \centering
\includegraphics[width=0.95\columnwidth]{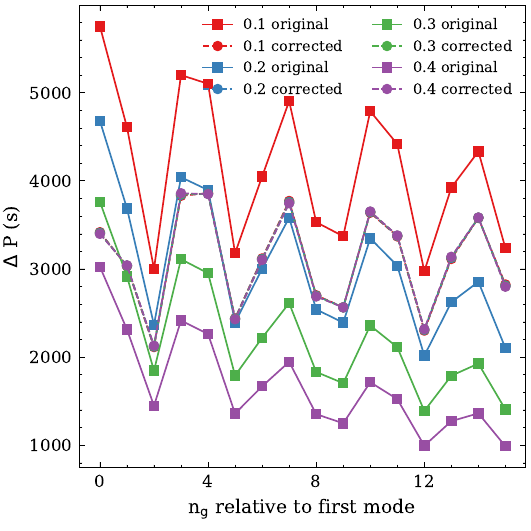}
\caption{Sketch of correction that we apply to a period spacing pattern due to rotation.} 
\label{fig_rotation_correction}
\end{figure}

Figure \ref{fig_rotation_correction} sketches the impact of the correction we applied to period spacing patterns to account for rotation. The four solid lines show period spacing patterns for models with the same stellar structure, but with different $\Omega/\Omega_{\rm crit}$ to compute the modes. The values of $\Omega/\Omega_{\rm crit}$ are 0.1, 0.2, 0.3, and 0.4 for the red, blue, green, and purple lines, respectively. The four dashed lines (all overplotted on each other) show the period spacing pattern after application of our correction. Our correction is achieved by fitting the period spacings with an inverse cubic function as a function of mode number, dividing the period spacings by this fit, and then scaling by a constant factor of 3000. The goal of the correction is to obtain a pseudo period spacing pattern that depends only on the stellar structure and not on the rotation. Fig. \ref{fig_rotation_correction} demonstrates that our correction achieves this goal, as the four models same stellar structure and different rotations are almost identical after the correction is applied.

\section{Dependence of seismic constraints on structure parameters} \label{app_constraints_params}
\begin{figure}[h] \centering
\includegraphics[width=0.95\columnwidth]{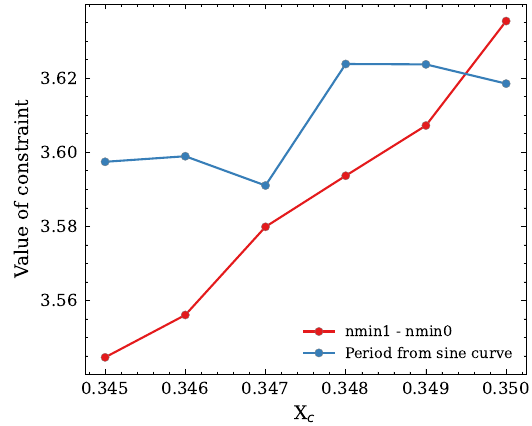}
\caption{Dependence of the period obtained by fitting a sine curve through a period spacing pattern for models with different $X_c$  values (blue line) compared with the value obtained from our choice of constraint to find the minima of the dips described in Sect. \ref{sec_constraints} (red line). } 
\label{constraint_dependence_appendix}
\end{figure}

In order for the our Jacobian procedure to converge towards the correct solution, the seismic constraints must depend monotonically on the structure parameters, and the point at which the value of a given constraint in the target and the model are equal  should roughly correspond to when the value of the structure parameter is correct.
Figure \ref{constraint_dependence_appendix} compares two possible constraints related to the periodicity of the dips in the period spacing pattern. The first is the period obtained by fitting a sine curve through a period spacing pattern and the second is the difference between two minima defined by our Eq. \ref{eqn_min}. Figure \ref{constraint_dependence_appendix} plots the values of these constraints for six different models with the same stellar structure and rotation, except for the value of $X_c$, which varies from 0.345 to 0.350. The variation of $X_c$ is simply a way to vary the periodicity of the dips in the period spacing pattern for the purposes of this test. We see that the difference between two minima defined by our Eq. \ref{eqn_min} produces a monotonic variation as a function of $X_c$; however, the period from the sine curve does not. We used these kinds of tests to search for suitable seismic constraints to use in our Jacobian, and to understand how they impact the structure parameters $s_j$.

\section{Convergence for the model inside the parameter space}
\begin{figure*}[h] \centering
\includegraphics[width=0.85\textwidth]{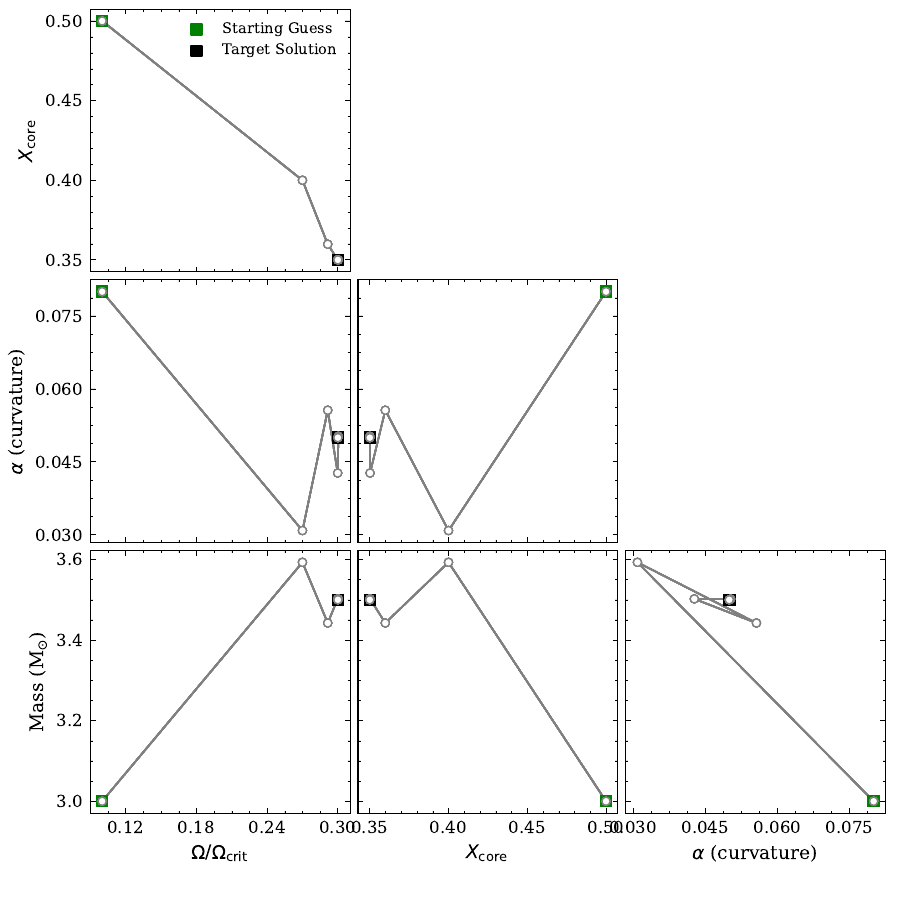}
\caption{Convergence through the parameter space for a model inside the parameter space, also presented in Fig. \ref{conv_inside_param_space} and Table \ref{table_conv_in}. The starting guess and target solutions are indicated by green and black boxes, respectively. Each iteration is indicated by a white point.} 
\label{conv_inside_param_space_paramplot}
\end{figure*}

To illustrate how the inversion procedure moves through the parameter space, Fig. \ref{conv_inside_param_space_paramplot} shows the evolution of the parameters $s_j$ for the same model as presented in Fig. \ref{conv_inside_param_space}.

\section{Other convergence examples for models inside the parameter space}
Table \ref{conv_out2} shows other examples of the convergence of models inside the parameter space with artificial data. The values for $s_j$ of the artificial target are $M = 3.5\msun$, $X_c = 0.35$, $\Omega/\Omega_{\rm crit} = 0.30$ and $\alpha = 0.05$. Models A - E converge to the correct solution. Model F is an example of a divergence that occurred because the initial guess was too far away from the target solution

\begin{table*}[h!]
\centering
\caption{Other examples of the convergence of models inside the parameter space with artificial data.}
{\footnotesize
\begin{tabular}{l rrrrrrrr}
 & \multicolumn{4}{c}{{\bf Starting $s_j$}} &  \multicolumn{4}{c}{{\bf Final $s_j$}} \\
 {\bf Model} &{\bf Mass (M$\bm{_{\odot}}$)} & $\bm{\Omega/\Omega_{\rm crit}}$ & $\bm{X_c}$ & $\bm{\alpha}$ & {\bf Mass (M$\bm{_{\odot}}$)} & $\bm{\Omega/\Omega_{\rm crit}}$ & $\bm{X_c}$ & $\bm{\alpha}$ \\  \midrule
A                    & 3.8000       & 0.2700       & 0.3000       & 0.0200       & 3.5000       & 0.3000       & 0.3500       & 0.0500       \\ \midrule
B                    & 3.8000       & 0.3500       & 0.5000       & 0.0200       & 3.4999       & 0.3000       & 0.3500       & 0.0600       \\ \midrule
C                    & 3.0000       & 0.1000       & 0.5000       & 0.0800       & 3.5000       & 0.3000       & 0.3500       & 0.0500       \\ \midrule
D                    & 3.2000       & 0.1000       & 0.3000       & 0.0300       & 3.5000       & 0.3000       & 0.3500       & 0.0500       \\ \midrule
E                    & 4.0000       & 0.4000       & 0.2500       & 0.0100       & 3.5000       & 0.3000       & 0.3500       & 0.0500       \\ \midrule
F                    & 3.8000       & 0.2700       & 0.6000       & 0.0200       & 2.2000       & 0.1005       & 0.6700       & 0.0100       \\ \midrule
\label{conv_out2}
\end{tabular}}
\end{table*}

\section{Convergence for models outside the parameter space}
\begin{figure*}[h] \centering
\includegraphics[width=0.85\textwidth]{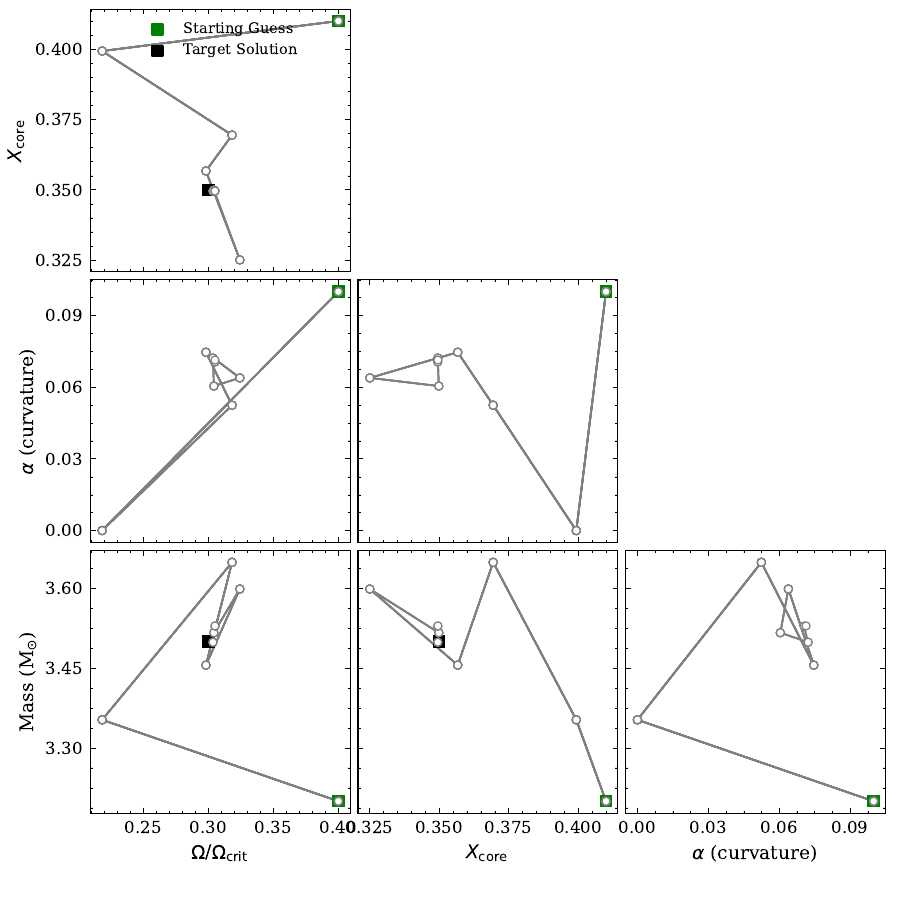}
\caption{Convergence through the parameter space for a model outside the parameter space, also presented in Fig. \ref{conv_outside_param_space} and Table \ref{table_conv_out}. The starting guess and target solutions are indicated by green and black boxes, respectively. Each iteration is indicated by a white round point. Note that since this model is outside our parameter space, the curvature $\alpha$ is not well defined.} 
\label{conv_outside_param_space_paramplot}
\end{figure*}

To illustrate how the inversion procedure moves through the parameter space, Fig. \ref{conv_outside_param_space_paramplot} shows the evolution of the parameters $s_j$ for the same model as presented in Fig. \ref{conv_outside_param_space}.

\section{Examples of degeneracy between the stellar structure and the period spacing patterns}
Selected models that exhibit some examples of degeneracy between the period spacing patterns and the stellar structure.

\begin{figure*}[h] \centering
\includegraphics[width=1\textwidth]{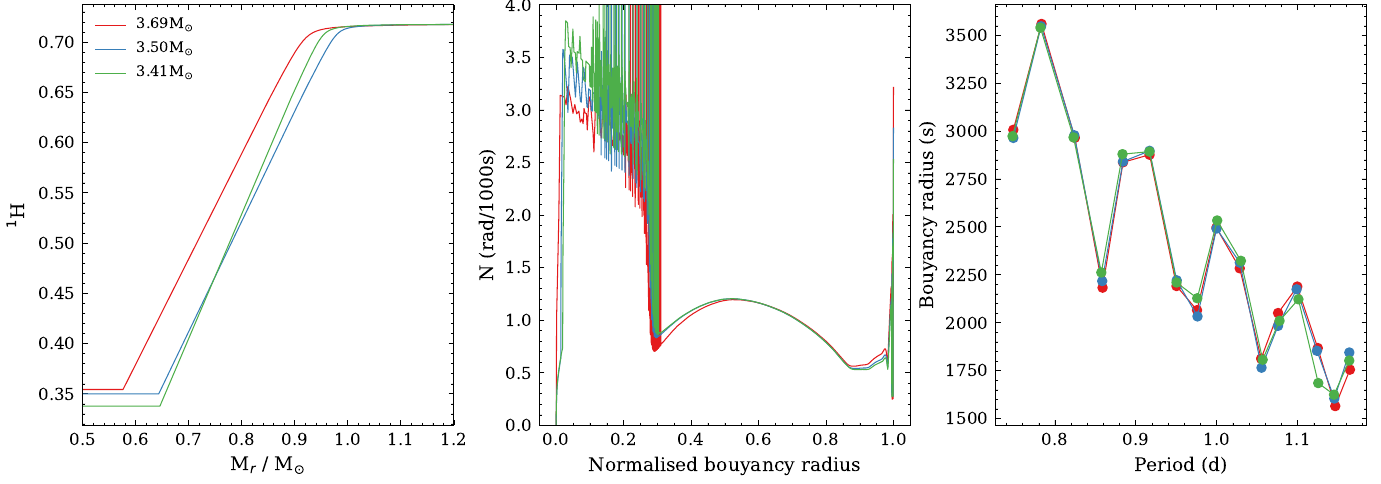}
\caption{Example of degeneracy: three different stellar structure models with similar period spacing patterns. All three models have the same values for the metallicity $Z$, $X_s$ and mode classification. They were constructed by setting the values of $M_{\mu}$ and $H_{\rm slope}$ and allowing the values of mass $M$, $X_c$, $\Omega/\Omega_{\mathrm{crit}}$ and curvature $\alpha$ to vary. The red, blue and green models correspond to $M_{\mu}$ = 0.25, 0.25, and 0.28 and $H_{\rm slope}$ = 3.8, 3.8, and 4.2, respectively. These changes are reflected in the left panel. {\it Left panel}: Hydrogen abundance profile as a function of mass coordinate for each model. {\it Middle panel}: \bv profile as a function of normalised buoyancy radius. {\it Right panel}: Period spacing patterns. } 
\label{degeneracy_example}
\end{figure*}

\begin{figure*}[h] \centering
\includegraphics[width=1\textwidth]{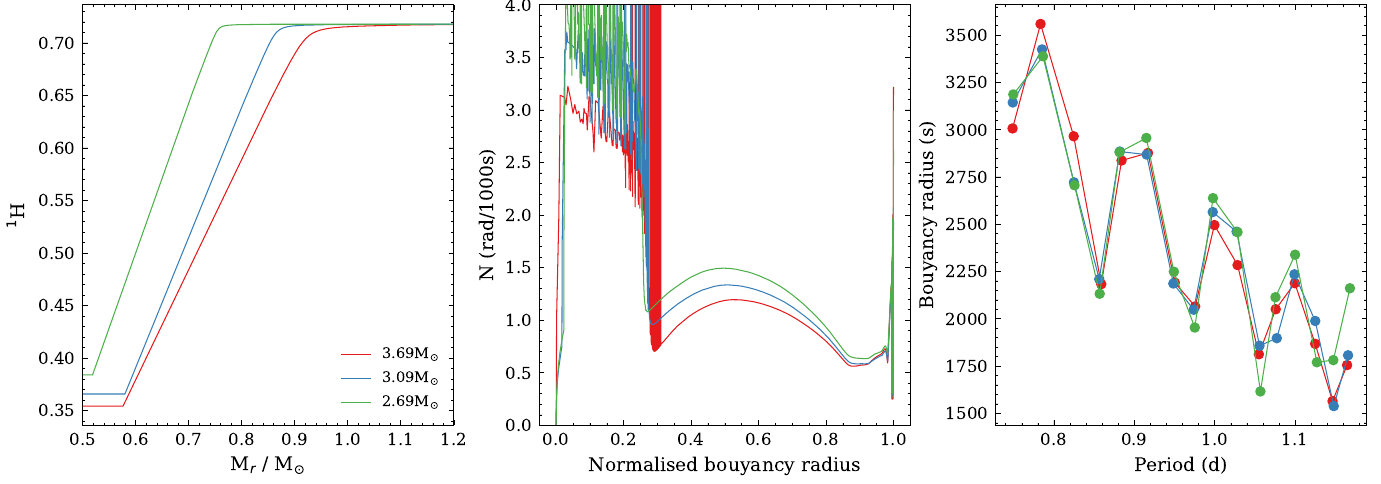}
\caption{Example of degeneracy: three different stellar structure models with similar period spacing patterns. All three models have the same values for $H_{\rm slope}$, metallicity $Z$, $X_s$. They differ in their values of mass $M$, $X_c$, $\Omega/\Omega_{\mathrm{crit}}$, $M_{\mu}$ and curvature $\alpha$ and mode classification (the lower order modes plotted in the period spacing pattern have radial orders n = 14, 15, and 16, respectively, for the red, blue, and green lines). Quantities plotted are the same as Fig. \ref{degeneracy_example}} 
\label{degeneracy_example_mode}
\end{figure*}

\bibliographystyle{aa}
\bibliography{refs.bib}

\end{document}